# Enhanced Surface Photon Drag on Plasmonic Metamaterials: A Fizeau-Doppler Shift Study

Leilei Yin and Nicholas X. Fang

Department of Mechanical Science and Engineering and the Beckman Institute of Advanced Science and Technology, University of Illinois, Urbana 61801, Illinois

Abstract

In this work, we numerically study the Fizeau-Doppler shift of surface modes in a moving plasmonic metamaterial. At proper surface plasmon resonance conditions, a slow group velocity is resulted in the metamaterials, and our simulation indicates a phase shift up to 4 orders of magnitude larger than the normal mode. In addition, both appreciable positive and negative photon drag could be observed. We also found an abrupt transition in the sign of phase shift when silver film reaches a critical thickness. This opens new opportunities in applications such as compact optical gyroscopes and accelerometers.

A light beam incident upon a moving dielectric medium is, in general, subject to a frequency shift due to a photon drag known as Fizeau-Doppler effect [1]. The study on Fizeau-Doppler effect has a long history and controversy [2,3]. One reason is that the Fizeau effect in a normal medium of linear dispersion is generally weak and experimentally difficult to detect. In 1970s, anomalous Fizeau-Doppler drag effect was predicted in dispersive medium with slow or even negative group velocities [4, 5]. However in the past, such effect could only be demonstrated with the presence of substantial absorption [6-8]. Recent discovery of metamaterials [9, 10] have opened an exciting gateway to engineer anomalous light propagation, and even leading to negative refraction of light. Pioneering studies of photonic crystals undergoing rigid translation and rotation[11, 12] are excellent examples of unique photon drag effect in these engineered metamaterials.

In this work, we propose to enhance the effect of photon drag through plasmonic metamaterials. Our motivation is based on the strong dispersion of surface modes and substantial energy exchange during the photon-polariton transition. A recently reported research on Goos-Hänchen shift [13, 14] indicates that optimized surface plasmon resonance may lead to great enhancement for Fizeau-Doppler shift.

To excite the surface plasmon resonance, our model system utilize a film of metamaterial with permittivity ($\varepsilon_1$) of varying thicknesses deposited on a high index medium of permittivity ($\varepsilon_0$) as shown in Fig. 1(a), which is known as the Kretschmann-

Reather attenuated total reflection (ATR) device. Here $\varepsilon_j$ is the permittivity of the different media of the prism, the metal layer and air ($\varepsilon_2$). At the surface plasmon resonance, the incident electromagnetic field decays exponentially in the film and excites the surface plasmon wave propagating along the metamaterial-air interface. The reflection coefficient in the ATR configuration, derived from the Fresnel's equation for a three-layer system, is $R = \dfrac{r_{01} + r_{12} \exp(2i\delta)}{1 + r_{01} r_{12} \exp(2i\delta)}$ [15], where $r_{ij}^{TM} = \dfrac{\varepsilon_j k_{zi} - \varepsilon_i k_{zj}}{\varepsilon_j k_{zi} + \varepsilon_i k_{zj}}$ and $r_{ij}^{TE} = \dfrac{k_{zi} - k_{zj}}{k_{zi} + k_{zj}}$ are the reflection coefficients of TM and TE electric field on interfaces. $\delta = -k_{z1} d_1$ represents the complex phase shift during the twice wave propagation through the thin film. Be noticed that the reflection coefficients of electric field are dependent on the polarization states. Only TM mode electric field is responsible for exciting the surface plasmons. In the ATR configuration $k_{z1}$ and $k_{z2}$ are in evanescent mode, and since the metal film is typically very thin (tens of nanometers), the absorption of $k_z$ mode is neglected.

If the ATR device undergoes a rigid translation parallel to the interface, the frequency and wave vector of electromagnetic wave in the moving frame $\Sigma'$ are corrected by one-dimensional Lorentz transformation equation $\omega' = \gamma(\omega - k_x V)$ and $k_x' = \gamma\left(k_x - \dfrac{V\omega}{c^2}\right)$, where $\gamma = (1 - V^2/c^2)^{-1/2}$ and $V$ is the speed of translation. In the $\Sigma'$ frame, the dispersion relation inside a medium is $\varepsilon(\omega')\omega'^2 = c^2(k_x'^2 + k_z'^2)$, similar to it as in rest $\Sigma$ frame. If $V \ll c$, the linearized Lorentz transform can be used. By inserting the relativistic corrected wave vector and frequency, $\omega' = \omega - k_x V$ and $k_x' = k_x - \dfrac{V\omega}{c^2}$, we can obtain the dispersion curve of SPP in the moving frame, as shown in Figure 1(a). In comparison to the shift $\Delta k_{light}$ of light line, the change in wavevector of SPP at the same frequency is more pronounced as $\Delta k_{SPP}$ shown in Figure 1(a). Figure 1(b) also shows a proposed configuration of ATR method allows continuous motion of the structure while maintaining resonant condition of SPP. For simplicity, the drag effect of light beam in the passage of glass tube is not considered.

Let's focus on one side of the setup. The physical variable experiencing relativistic change during the rotational motion is $k_z'$, which gives the additional phase shift due to specific relativity:

$$k_z' = \left[ \frac{\varepsilon(\omega)}{c^2}(\omega - k_x V)^2 - \left(k_x - \frac{\omega V}{c^2}\right)^2 \right]^{1/2} \qquad (1)$$

By replacing $k_z$ with $k_z'$, we acquire complex reflection coefficients for both TM and TE mode in the ATR configuration undergoing a rigid translation. The absolute values of the complex coefficients describe the amplitude of reflection while the angles determine the phase shifts. By subtracting the phase shifts of reflection coefficients with and without rigid translation of the ATR setup, we are able to observe the additional phase shift induced by the Fizeau-Doppler drag. Figure 2 shows the calculation results from Fresnel equation of three-layer system with various thickness of a thin metamaterial film deposited on BK7 glass prism. In this case, we assume the wavelength of light in vacuum is set at 543.5 nm, and the thin metamaterial film is taken as silver. The permittivity of BK7 glass is $\varepsilon_0 = 2.307$, the permittivity of silver used in this calculation is $\varepsilon_1 = -10.5459 + i0.8385$ [16] and the permittivity of air is at unity ($\varepsilon_2=1$). We first assume the translation speed of the ATR setup at $V = 10^{-8} \times c = 3$ m/s in our calculation. This translation speed is small enough for us to neglect the higher order relativistic correction of $\gamma = \sqrt{1-(V/c)^2}$ and the dispersion of $\varepsilon_0(\omega')$ and $\varepsilon_1(\omega')$. The calculation reveals that an additional phase shift caused by Fizeau-Doppler drag has an angular resonant peak corresponding to the resonant excitation of surface plasmons in the ATR method. When the thickness of silver film reduces, the resonant angle has a slow shift toward smaller incident angle, a similar phenomenon as the resonant excitation of surface plasmons on silver thin film [14]. However, the magnitude the Fizeau-Doppler phase shift associated with the surface plasmon excitation as related to the incident angle is a striking effect compared to the amplitude of reflection in an ATR method. The Fizeau-Doppler phase shift increases from $-1.12 \times 10^{-7}$ radian to almost $-1.20 \times 10^{-6}$ radian, over 10 fold, when the thickness of silver film increases from 30 nm to 45 nm. When thickness further increases from 45 nm to 50 nm, the Fizeau-Doppler phase shift shows a dramatic transition from $-1.20 \times 10^{-6}$ radian to $+1.02 \times 10^{-6}$ radian. We attribute such transition to the dispersion of surface Plasmon modes in films of finite thicknesses [17]. In such thin films it is possible to excite degenerate polariton modes (when $1+ r_{01}r_{12} \exp(-2ik'_{z1}d_1)$ approaches zero) of different symmetry, thus providing different phase lag. Further increase of thickness of silver film begins weakening the phase shift but toward a smaller resonant incident angle. Calculation even shows a silver film with thickness around 47 nm can have narrower phase shift peak with amplitude a few orders higher than the curve of 45nm and 50 nm thick silver films. In the future experiments, to observe such narrow peaks of phase shift peak and more abrupt transition from negative and positive phase transition, it might be critical to tightly control roughness and interfacial scattering of the deposited silver films. This is recently achieved through the use of seed layers during silver deposition [18].

The Fizeau-Doppler drag is a physical phenomenon common to different polarization state of electromagnetic wave. In an ATR configuration, both TE and TM mode incidence shall be affected by the rigid motion of the entire setup. Our calculation confirms that both modes should be corrected with special relativity. Yet, they respond to the translation at significantly different magnitudes. Figure 2 also shows the Fizeau-Doppler phase shift in the same conditions of wavelength and translation, but with different polarization states on 45 nm and 50 nm thick silver films. The TE mode Fizeau-Doppler phase shift is amplified by 1000 times to facilitate visual comparison with the phase shift of TM mode. In the TE mode, the phase shift shows a nearly linear angular dependence in a narrow span of incident angles. This is mainly due to the reflection coefficient $r_{ij}^{TE}$ on each interface has a much simpler relation of incident angle, without involving surface plasmon excitation. Figure 2 clearly shows that around the surface plasmon resonance on 45 nm and 50 nm thick silver films of TM mode the Fizeau-Doppler phase shift has an enhanced factor of over three orders of magnitude compared to the TE mode in otherwise the same conditions.

If we keep the incident angle to the values that give the maximum Fizeau-Doppler phase shift in TM mode and vary the translation speed, we then are able to calculate the phase shift rate of a certain ATR setup. In our study, we varied the speed of rigid translation from 0.1 m/s to 30 m/s on both 45 nm and 50 nm thick silver films. The incident angle of peak phase shift at various translation speeds is 43.93° on 45 nm thick silver film and 43.89° on 50 nm thick silver film. The Fizeau-Doppler shift as a function of translation speed of TM and TE modes on both films shows excellent linearity as a function of the translation speed. The calculation reveals that on 45 nm silver film the phase shift of TM mode and TE mode are $-3.978 \times 10^{-7}$ radian/m/s and $1.147 \times 10^{-10}$ radian/m/s, and on 50 nm silver film $3.402 \times 10^{-7}$ radian/m/s and $3.620 \times 10^{-11}$ radian/m/s respectively. The small magnitude of phase shift in the TE mode clearly indicates the difficulty of observing Fizeau-Doppler effect in linear medium, and usually requires a very long light path for the detection. Yet in contrast, on 50 nm silver film the TM mode Fizeau-Doppler shift shows almost 4 orders enhancement over TE mode contributed by the surface plasmon resonance in the ATR method. On 45 nm thick silver film, the enhancement factor is also over 3000.

    Not only the thin film structure contributes to the dispersion of SPP wave, metal itself presents dispersive dielectric coefficients over a wide spectrum of light. This dispersion inevitably affects SPP enhanced Fizeau-Doppler shift too. Using the same mathematic equations with frequency-dependent dielectric coefficients of the media, one can conveniently study the Fizeau-Doppler shift of various wavelengths taken from Reference 15. Figure 3 shows the Fizeau-Doppler shift of 6 selected wavelengths on 50 nm thick silver film. With 50 nm thickness, Fizeau-Doppler shifts of all selected wavelength are positive with different magnitudes and resonant angles. The widths and magnitudes of each spike are compared in Figure 3 too. We can see for each specific thickness of silver film, there is an optimal wavelength with higher enhancement factor of phase shift. This simply requires a divergence of reflectivity in the moving frame, or

$1 + r_{01}r_{12}\exp(-2ik'_{z1}d_1)$ diminishes, a necessary condition for surface plasmon excitation on the thin film. Likewise, optimal enhancement of such shift on a specific wavelength can be achieved with a carefully designed film thickness. Notice that the width of the sharpest spike, though greatly enhanced in magnitude, at 590 nm is merely $0.05°$, about 10% of that of a SPP resonance angle [15]. Such narrow angular width could pose a challenge to the experiments on the spectral characteristics of laser source and precision of optomechanical apparatus.

In summary, by adding additional relativistic correction terms into Fresnel's equation for an ATR configuration, we numerically studied the Fresnel-Fizeau drag involving resonant excitation of surface plasmons on thin silver films with various thicknesses. We find that the Fizeau-Doppler phase shift shows a giant enhancement (3-4 orders) factor, characterized by the resonant excitation of surface plasmons on the thin silver films. Such maximum phase shift occurs at the critical angle of resonant excitation of surface plasmons. Furthermore, the Fizeau-Doppler phase shift makes an abnormal and abrupt transition from negative to positive when the thickness of silver increases over a critical value, in our case, around 47 nm for 543.5 nm light.

The classical Fizeau shift is known for its weak signal. However our study on the Fizeau-Doppler phase shift through plasmonic metamaterials indicates that the phenomenon can be greatly enhanced by simple modifications to the experimental apparatus. While this study presented the enhanced Fizeau-Doppler effect on silver film, it is generally applicable to a broad range of metamaterials from optical to microwave frequencies. With the development of low-loss and frequency tunable metamaterials, we expect this discovery to open new avenues in applications such as compact accelerometers or optical gyroscopes.

The authors are grateful for the financial supports from the Defense Advanced Research Projects Agency (grant HR0011-05-3-0002) and Office of Naval Research (grant N00173-07-G013).

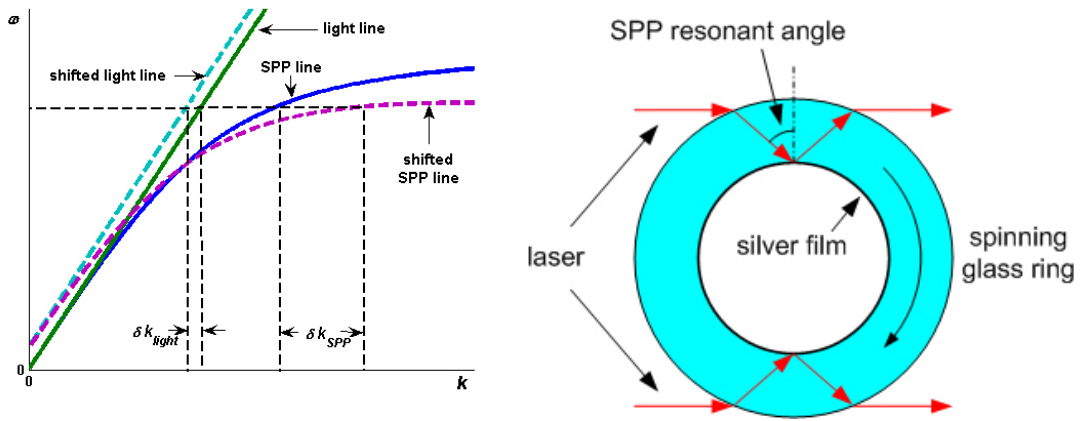

Figure 1. Left: The SPP dispersion curve on a semi-infinite interface in stationary and motion frames. The drag of metal surface causes light line shift and the dispersion curve of SPP complicated transition. Right: A modified configuration of the ATR method for Fizeau-Doppler shift experiment that enables sustained translation of media. The upper and lower laser beams sustain opposite Fizeau-Doppler shifts of SPP waves but identical drags of glass ring. Interference of the two beams should present solely the FD shift of SPP wave.

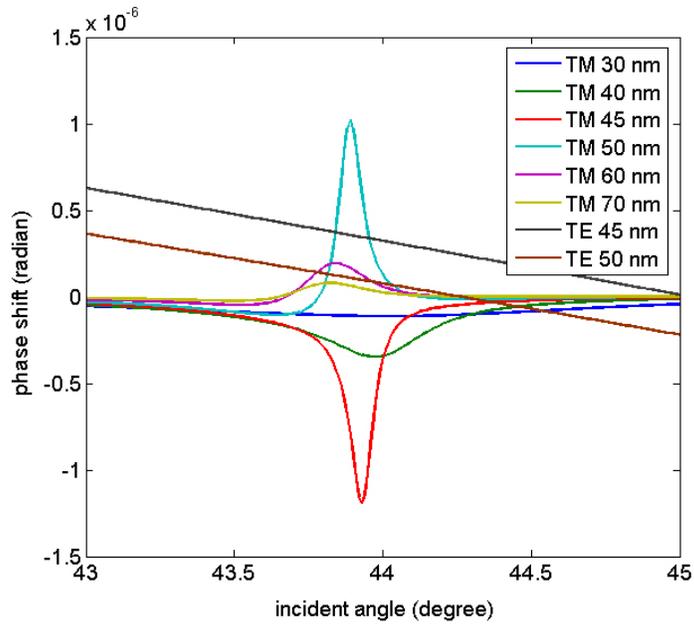

Figure 2. The Fizeau-Doppler phase shifts in an ATR setup with incident light around surface plasmon resonant angle. The speed of translation is 3 m/s, and the wavelength of light in vacuum is 543.5 nm. Inset indicates the thickness of silver film and polarization modes used in calculation, the amplitude of TE mode shifts is multiplied by 1000 to be visually comparable to TM mode shifts.

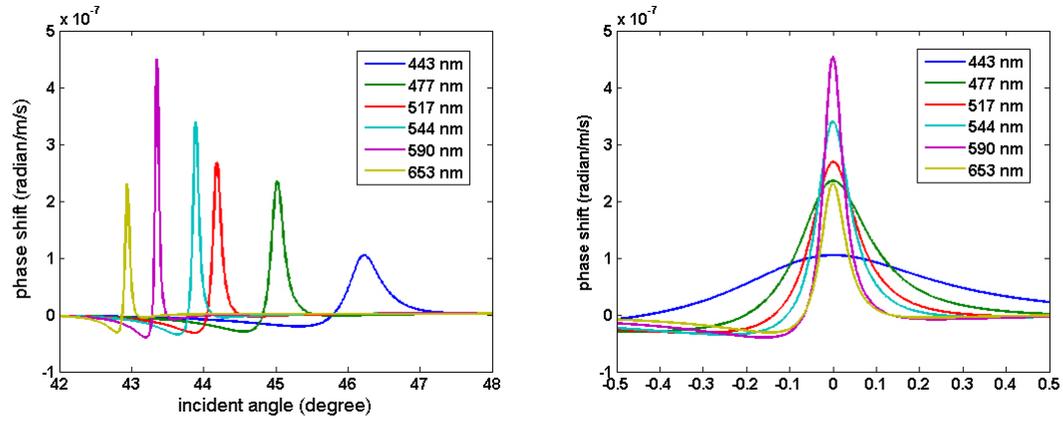

Figure 3. Left: Fizeau-Doppler frequency shift per unit speed on 50 nm thick Ag film with various wavelengths. Right: Incident angular width of shift peaks of each wavelength. The narrowest peak of 590 nm wavelength has a full width of mere 0.05° at half height, considerably narrower than the SPP resonance angle width around 0.5°.